\documentclass[12pt]{article}
\usepackage[english]{babel}
 \usepackage[square,numbers,sort&compress]{natbib}
\usepackage{amsmath}
\usepackage{amssymb}
\usepackage{spalign}
\usepackage{graphicx}
\usepackage[labelfont=bf]{caption}
\usepackage{multirow}
\usepackage{booktabs}
\usepackage{authblk}
\usepackage{xcolor}

\author[1,2]{A. Lattanzi\footnote{Corresponding author: ambra.lattanzi@ifj.edu.pl \hspace{0.2cm} ambra.lattanzi@gmail.com}}
\author[2]{G. Dattoli}
\author[3]{G. Baldacchini}
\affil[1]{H. Niewodnicza\'nski Institute of Nuclear Physics Polish Academy of Science IFJ-PAN, Krak\'ow, Poland}
\affil[2]{ENEA Research Center of Frascati, Frascati, Rome, Italy}
\affil[3]{GBConsigli, via Guglielmo Quattrucci 246, Grottaferrata, Rome, Italy}

\date{}                     
\begin{document}
 \title{Physics and Mathematics of the Photoluminescence of Complex Systems}
\maketitle
\vspace{-2cm}
\begin{abstract}
The photoluminescence (PL) of thermally evaporated Alq3 thin films has been studied in a few samples annealed and non-annealed and afterwards exposed to the laboratory atmosphere for over six years. It was found that the measured emission intensity decays with a long lifespan and with four different time-spectral behaviors, which imply the existence of four molecular aggregations, or components. In particular, the time behavior of each component follows the trend of a Kohlrausch$-$Williams$-$Watts (KWW) function, which is well known in mathematics but without any physical meaning. Here, by introducing the concept of the material clock, the system has been described by a damped harmonic oscillator, which in certain conditions, fulfilled in the present case, allows the expansion of the KWW function in the so-called Prony series. The terms of this series can be attributed to chemical and physical processes that really contribute to the decay, i.e. the degradation, of the Alq3 thin films when interacting with internal and environmental agents. These insights unveiled the usefulness of proper mathematical procedures and properties, such as the monotonicity and the complete monotonicity, for investigating the PL of this ubiquitous organometallic molecule, which possesses one among the highest emission yield. Moreover, this method is also promising for describing the photoluminescent processes of similar organic molecules important both for basic research and optoelectronic applications.
\end{abstract}
\newpage

\section{Introduction}\label{sez1}

The Kohlrausch$-$Williams$-$Watts (KWW) function is widely used to describe the non-exponential relaxation dynamics of complex systems that has been and still is a topic of intense research \cite{c1,c2,c3,c4,b,c5,c6,c7,c8,c9,c10,c11,c12,c13,c14,kww04,c15,c16,c17,c18,c19,c20,c21,ngai,c22,c23,c24,c25,c26}. This function has the form \cite{kww01,kww02,kww03}
\begin{equation}\label{1}
f(t)=e^{-\Big(\frac{t}{\tau}\Big)^{\beta}},
\end{equation}
where the characteristic relaxation time $\tau$ and the KWW exponential $\beta$ are positive real parameters. Lastly $t\in[0,\infty)$ denotes the time variable. The range of the KWW exponential $\beta$ can be divided into two parts where the case $\beta=1$ represents a watershed. For $0<\beta\leq1$, the KWW function \eqref{1} is a stretched exponential function, otherwise (i.e. for $\beta>1$) the relaxation function \eqref{1} is named compressed exponential. Their names describe the time-dependent behaviors with respect to the pure exponential. If the function \eqref{1} is stretched, its time-dependent behavior is slower than the exponential function and it is widely used to model relaxation in glassy systems \cite{c14,c22,c23,c24,c25}. On the other hand, the compressed case indicates a faster than exponential trend and it is commonly used to model the out-of-equilibrium soft materials \cite{c7,c8,c9,c13,c18,c19,c20}. The ubiquitous nature of the KWW function and its success in modelling experimental results do not imply that {the relevant role is not straightforwardly understood within an independent  non phenomenological model. The singularity in the first derivative for values of $\beta$ in the interval $(0,1)$ is an element of serious concern which raises questions on the nature of the variables $t$ used to define it in a specific physical context.}\\
\\
The aim of this paper is to introduce a phenomenological model able to unveil the meaning of the KWW function and the nature of its singularity resorting to a comprehensive approach in agreement with the experimental data and able to define the role of the monotonicity in the mathematical modelling of relaxation processes.\\ This model and its {appropriate} physical interpretation pave the way to a novel approach in the analysis of the time-resolved photoluminescence. It explains anomalous behavior and frame the results known in literature (see for example \cite{c1,c2,c3,c4}) in a different {fashion}. {This point of view, developed in the article, allows the understanding of issues linked to the function itself }(in its stretched and compressed versions) but also to the realization of a general model where the role of the restoring and frictional forces have been related to a physical quantity as the time-dependent reduced mass which is further linked to the total decay rate and ultimately, to the hidden dynamics of the complex system.  To understand the crucial value of the model and the approach here presented, we should remind the difficulties in the interpretation of the nonlinear friction mentioned in \cite{lukichev}. The physical approximation to a first order differential equation elegantly circumvents the mathematical obstacle in the solution of a second-order differential equation with time-dependent coefficients but the price to be paid is in the understanding of the nature of the nonlinear forces. The strategy here presented preserve the second order differential equation and introduce the nonlinear time variable as a \textit{change in the coordinates}. After obtaining the solution, we can restore the usual (linear) time and the interpretation of the nonlinear forces are embeds in its time-dependent coefficients. The second order differential equation now describes a damped harmonic oscillator with time-dependent mass and frequency and it can outline the dissipation of the complex system analyzed. \\ Among the complex systems described by the KWW function, we will consider as a test-case the forerunner of photoluminescent materials, Tris-(8-hydroyquino- line)aluminum molecule, or more briefly Alq3 (see Fig. \ref{fig1}). 
\begin{center}
\includegraphics[scale=0.2]{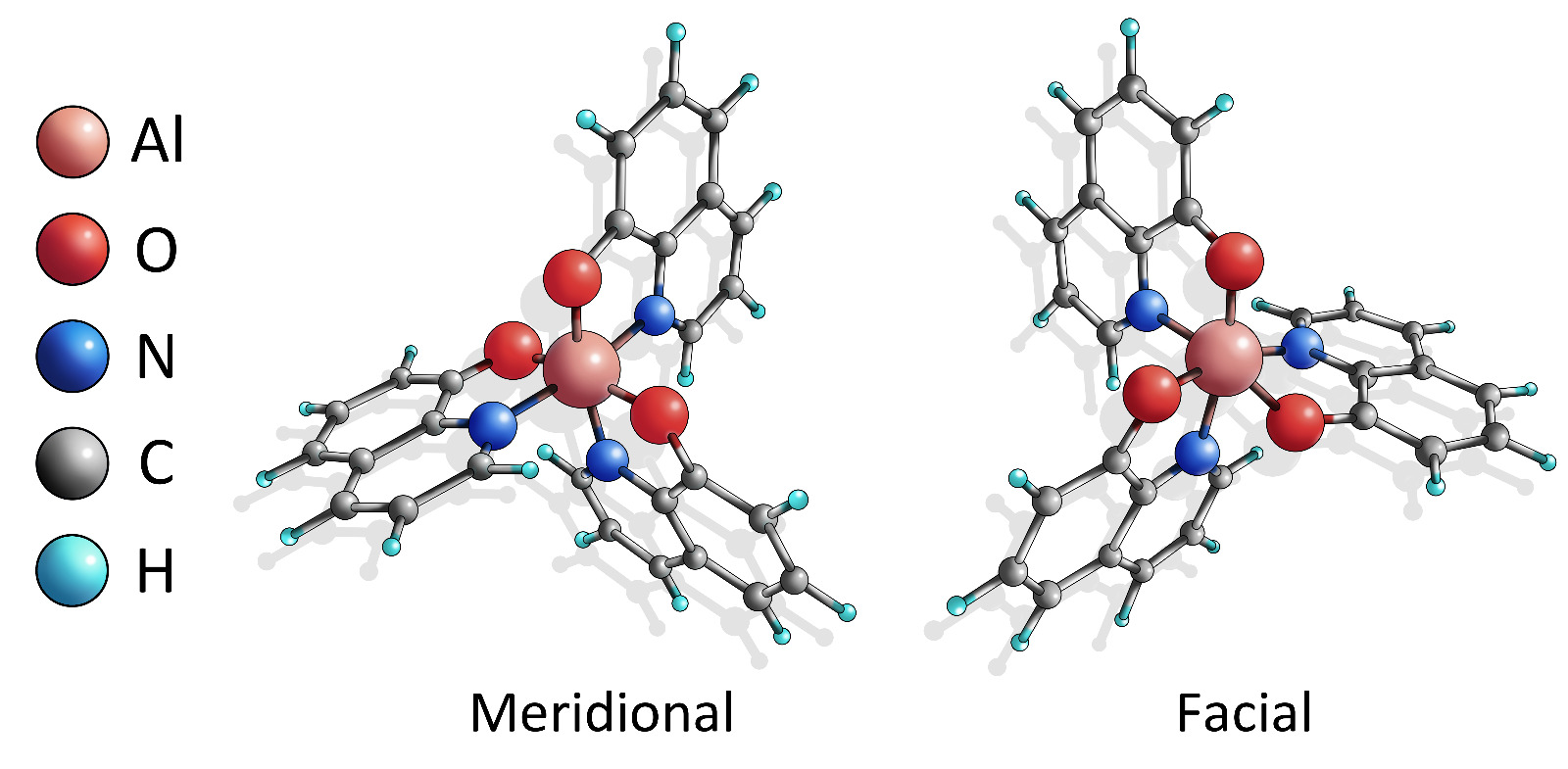}\label{fig1}
\captionof{figure}{Tris-(8-hydroyquinoline)aluminum molecule (Alq3). The formula of the chemical compound is $Al(C_9H_6NO)_3$ and it is illustrated in both configurations.}
\end{center}
The choice of the Alq3 molecule relies on two facts: first and foremost, its practical and historical importance in rampant photonic technological applications as the Organic Light Emitting Diodes (OLEDs) \cite{c27,c28,c29,c30,c31,c32} and secondly, the huge dataset collected in recent literature \cite{c1,c2,c3,c4}. Indeed, the study of photoluminescence (PL) of thin films of Alq3 represents a relevant and unique case in the history of solid-state matter, both for the different environmental conditions utilized and the extremely wide time range considered, more than six years of measurements!

The paper is divided according to the main steps of the proposed model. {\bf Section 2} focuses on the physical origin of the KWW function. This function comes out of an initial value problem that involves a second-order differential equation, where the photoluminescent system is modelled as a damped harmonic oscillator. The problem has been addressed in a system of reference different from the laboratory one, named material frame of reference. This change of perspective promoted a simplification from a mathematical point of view in favor of a clearer physical interpretation not only for the relaxation function but also for its singularity. {Last but not least, it restores the symmetry from a dynamical point of view between the dispersion and the relaxation models that are both described by a damped harmonic oscillator.} In {\bf Section 3}, the KWW function is expanded in Prony series so that it is possible to interpret the terms involved in a physicol-chemical context. In fact, the series expansion strategy allows to highlight how the PL intensity emission is the sum of processes that the KWW function elegantly summarizes. {The approximation keeps into account only the zero-th and first-order terms so it will be possible to bridge the known results in literature defined on a bi-exponential model with  the one presented in this work}. In {\bf Section 4}, the results and the considerations obtained in the previous Section are applied to an ideal photoexcited state in order to fix the idea before applying the model to a real physical system. {\bf Section 5} highlights the importance of the role of monotonicity and complete monotonicity in modelling relaxation processes. {\bf Section 6} is devoted to present a test-case based on the data collected on the photoluminescence (PL) of the Alq3 molecule. This Section illustrates how the approach can fit experimental data in two different situations in order to make evident the peculiar features in the time-resolved PL. In {\bf Section 7} we validate the approach illustrating how the pumping and damping mechanism rules the dynamics in photoluminescent system. The universality of the approach can be extended also to other photoluminescent materials. Finally, {\bf Section 8} is devoted to take a summary of the results obtained and some ideas as possible extensions of this research.


\section{The KWW function and its single and dual dynamics}\label{sez2}
The PL emission is the result of the relaxation of the luminophores. The term “luminophores” indicates the emitting centres of luminescence generated by photoexcitation of the ground state in atoms, molecules, and chemical compounds. As time passes, the number of luminophores that are still emitting begins to decrease. Specifically, the luminophores are quenched as the excitons (bound states of the photoexcited electron and its hole) relax. As a consequence the luminescent intensity decrease. Therefore, the intensity of the PL emission $I(t)$ can be modelled as a fraction between the number of the luminophores at a specific time $n(t)$ that are still emitting and the initial number of the same luminophores $n(0)=n_0$.

From a dynamical point of view, this degradation in the luminescent intensity is due to the frictional and restoring nature of the forces acting on the system. The radiative friction or resistance is the retarding force acting on charges when the electromagnetic radiation is emitted during a PL process. On the other hand, we have a restoring force due to the Coulomb attraction between the positive and negative charges that constitute the bound state in the excitons. Despite the simplified analysis of the forces acting on the system, the relaxation still exhibits complex and nonlinear behaviors that complicate the investigations on the material response and properties. We can simplify the problem changing the perspective with the introduction of a new reference system, called material reference frame, different from the classical laboratory one and centered on the material itself. The anomalous behaviors can be considered by introducing a flexible scale in the material frame of reference. Such a scale can stretch and compress the system giving an insight of what happens from the point of view of the material during the relaxation. According to this approach, during the relaxation process the material follows its own clock, called the material clock, which is in general different from the one set in the laboratory \cite{c33,c34,c35,c36,c37,c38}. The material clock represents the time measured by a clock whose rate itself changes: it can accelerate or decelerate. Consequently, in the laboratory the relaxation processes evolve with a different speed as the system \textit{ages}. The material clock is defined as 
\begin{equation}\label{2}
        t^*=t^\beta=\int_0^t \frac{\beta}{x^{1-\beta}}dx,
\end{equation}
where $\beta$ is parameter. If $\beta$ belongs to $0<\beta\leq1$, which from now on will be called the stretched range, the time flow is slower than the time in the laboratory. After crossing $t=1$, the time flow in the material starts to speed up and increase almost linearly. In this case, the material clock stretches the time scale. On the other hand, if $\beta>1$, the material clock compresses the time scale: the passing of time is faster at the beginning, and for $t\geq1$ it slows down its rate. This explains the name “compressed range”, when $\beta$ assumes values greater than 1. Taking into account the material frame of reference and the forces acting on the system, the relaxation dynamics can be described by the following initial value problem: 
\begin{equation}\label{3}
 \begin{split}
        &\frac{d^2n(t)}{d(t^{\beta})^2}+\frac{2\zeta}{\tau^\beta}\frac{dn(t)}{dt^\beta}+\frac{n(t)}{\tau^{2\beta}}=0\\
        &n(0)=n_0\\
        &\frac{dn(t)}{dt^\beta}\vline_{t=0}=-\frac{\zeta n_0}{\tau^\beta}.
    \end{split}
\end{equation}
	
The parameter $\zeta$ is the damping ratio and it critically determines the evolution of the system. The solution of \eqref{3} in terms of $\zeta$ reads:
\begin{equation}\label{4}
    n(t)=n_0 e^{-\zeta\Big(\frac{t}{\tau}\Big)^{\beta}}\mbox{cos}\Big(t^\beta\frac{\sqrt{1-\zeta^2}}{\tau^\beta}\Big).
\end{equation}
In the critically damped case ($\zeta=1$), the solution \eqref{4} is the Kohlrausch-Williams-Watts (KWW) function \eqref{1}. 

Considering the chosen material clock \eqref{2} and the derivative 
\begin{equation}\label{5}
    \frac{d}{dt^\beta}=\frac{1}{\beta t^{\beta-1}}\frac{d}{dt},
\end{equation}
we can formulate the initial value problem in the linear time reference frame, i.e. the laboratory and so, the relevant differential equation \eqref{3} writes
\begin{equation}\label{6}
\begin{split}
     & \frac{d^2n(t)}{dt^2}+\Big(\frac{1-\beta}{t}+\frac{2\beta t^{\beta-1}}{\tau^\beta}\Big)\frac{dn(t)}{dt}+\frac{\beta^2t^{2\beta-2}}{\tau^{2\beta}}n(t)=0\\
  &n(t)\vert_{t=0}=n_0\\
        &\frac{dn(t)}{dt}\vert_{t=0}=\spalignsys{0~~~\mbox{for $\beta>1$};-\frac{1}{\tau}~~~\mbox{for $\beta=1$};\infty~~\mbox{for $0<\beta<1$}.}
\end{split}
\end{equation}
We should focus the attention on the initial conditions. The first condition results to be the same of the initial value problem in the material reference frame:
\begin{equation*}
n(t^\beta)\vert_{t^\beta=0}=n(t)\vert_{t=0}=n_0,
\end{equation*}
as expected since the initial number of the luminophores should be the same in any reference frame. On the other hand, the second initial condition presents a variety of behaviors according to the value of the $\beta$ parameters. All these behaviors are a consequence of the choice of the reference frame and in order to support this idea we highlight how this initial condition in the laboratory reference frame is linked to the material reference frame. Mathematically, we have 
\begin{equation*}\label{velox}
\frac{dn(t)}{dt}=\frac{dn(t)}{dt^\beta}+\Big(\frac{1-\beta t^{\beta-1}}{\tau^\beta}\Big)n(t).
\end{equation*}
At $t=0$, it is possible to recover the initial conditions previously defined for the respective reference frames. The presence of this extra term supports our idea that the reference frame plays a key role in a clear understanding of the dynamics. To be sure that the singularity does not affect the physics, we analyze the nature of the singularity in the coefficients of the differential equation. \\
The coefficient in front of the first order derivative in \eqref{6} represents the time-dependent frictional term in the laboratory frame of reference and its form is a direct consequence of the material clock chosen \eqref{2}
\begin{equation*}
P(t):=\Big(\frac{1-\beta}{t}+\frac{2\beta t^{\beta-1}}{\tau^\beta}\Big),
\end{equation*}
whereas, the coefficient in front of the zero-th order term is
\begin{equation*}
    Q(t):=\frac{\beta^2t^{2\beta-2}}{\tau^{2\beta}}.
\end{equation*}
Both of these coefficients have a singularity for $t=0$ and it influences the derivatives of the KWW function and its initial condition in the laboratory frame of reference. In particular, the first derivative diverges at $t=0$ for $0<\beta<1$. However, the singularity is not particularly disturbing, in fact it is classified as a regular singularity \cite{bib23,bib90,bib108,bib154,bib155}. In other words, $P(t)$ diverges no more rapidly then $\frac{1}{t}$ and $Q(t)$ diverges no more rapidly than $\frac{1}{t^2}$. We remark that the singularity appears when we come back to the laboratory frame of reference and it does not exist in the material reference frame.\\ On the basis of all this information, we therefore suspect that the singularity in $t=0$ is a \textit{coordinate} singularity, i.e. a singularity that is nested in the choice of the reference frame.\\
The complex and anomalous behaviors, which we strategically embed in the material time, appear and are entirely enclosed in the time-dependent coefficients of the above differential equation but now they are no longer a problem but a crucial asset for interpreting the photoluminescence emitted by a real system. We focus our attention on the time-dependent frictional term $P(t)$ by searching for a related physical quantity, in order to unveil the mechanism behind the anomalous behavior.

Here we assume that the number of the luminophores plays the role of a generalized coordinate, considering that it is possible to define unequivocally how populated is the photoexcited state of the system. According to this, the problem can be formulated introducing the Lagrangian of an harmonic oscillator:
\begin{equation}\label{7}
    L=\frac{1}{2}\mu(t)\Big(\frac{dx(t)}{dt}\Big)^2-\frac{1}{2}\omega^2(t)\mu(t)x(t)^2, 
\end{equation}
where $\mu(t)$ and $\omega(t)$ denote respectively the time-dependent reduced mass and the time-dependent frequency of the system. Lastly $x(t)$ represents the generalized coordinate that defines uniquely the system. Applying the Lagrange equation, we can rewrite the differential equation \eqref{6} as:
\begin{equation}\label{8}
\begin{split}
\frac{d^2x(t)}{dt^2}+\frac{\dot\mu(t)}{\mu(t)}\frac{dx(t)}{dt}+{\omega^2(t)}x(t)=0,
\end{split}
\end{equation}
where $\dot{\mu}(t)$ is the time derivative of the reduced mass $\mu(t)$. 
Now we can compare \eqref{6} with \eqref{8}, and we get:
\begin{subequations}\label{9}
    \begin{align}\label{mueomega}
    \mu(t)&=mt^{1-\beta}e^{2\Big(\frac{t}{\tau}\Big)^{\beta}},
    \\
\omega(t)&=\frac{\beta t^{\beta-1}}{\tau^{\beta}}.
\end{align}
\end{subequations}
In conclusion, the Hamiltonian of the relaxation dynamics is given by
\begin{equation}\label{10}
    {\cal{H}}=t^{1-\beta}e^{2\Big(\frac{t}{\tau}\Big)^{\beta}}\Big[\frac{m}{2}\Big(\frac{dn(t)}{dt}\Big)^2+\frac{m n(t)}{2}\frac{\beta^2t^{2\beta-2}}{\tau^{2\beta}}\Big].
\end{equation}
Equation \eqref{10} generalizes the Caldirola-Kanai Hamiltonian \cite{c39,c40} where the time-dependence of the mass and frequency terms can be obtained from (9a) fixing $\beta=1$. {This result is particular intriguing since it paves the way to enhance the understanding of the role of quantum processes underlying this out-of-equilibrium system. After applying canonical quantization transformations, it is possible to move from classical to quantum mechanics using non-unitary transformations \cite{hpaper}. The natural emerging of the Hamiltonian \eqref{10} offers an overview of the subtle interplay between quantum and classical physical effects that characterize this \textit{mesoscopic} approach.\\ 
Since the quantum mechanical treatment of the problem is out of the scope of this paper, we should focus our attention on the information that can be obtained from the time-dependent reduced mass $\mu(t)$.} Its importance relies on the insights on the dynamics of the system. \\It can describe the presence of a damping and/or a pumping mechanism as it is linked to the inertia of the system and then it is proportional to the decay rate.
\begin{center}
\includegraphics[scale=0.7]{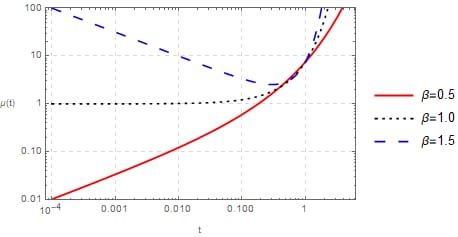}
\captionof{figure}{Reduced mass $\mu(t)$ as a function of time for $\tau=1$ and three values of $\beta$: $\beta=0.5$, $\beta=1$, $\beta=1.5$ where the decay is in the stretched, pure exponential and compressed range.}
    \label{fig2}
\end{center}
In the stretched range, when $\beta=0.5$, the reduced mass exhibits at least initially a fairly linear behavior. Then it starts to increase as rapidly as $\frac{t}{\tau}$ approaches the value $1$, see the solid red line in Fig. \ref{fig2}. As the reduced mass term $\mu(t)$ is proportional to the decay rate, the system presents a damping behavior since $\mu(t)$ is an always increasing function in the time domain.

The peculiar case of $\beta=1$ presented in Fig. \ref{fig2} shows that the reduced mass term $\mu(t)$ is almost constant in the first part of its evolution: the luminophores are quenched at the same rate and therefore, the surrounding environment does not have any influence on the dynamics of the system. However, as the luminophores emit radiation and then are quenched, the cloud of the virtual particles (i.e. phonons, photons, and polaritons inside the luminophores, and the reactive species in the atmosphere that surrounds and interact with the luminophores also through the excitons) deforms the initial situation creating micro-environments and configurations that cause a distribution of relaxation time and an increasing of the decay rate. The stretched KWW function describes a system of luminophores that from the very beginning are affected by the surrounding environment. These deformations can be physically modelled as traps where the excited particles are captured and then released. The key notion underlying the stretched KWW function is the single-nature of its dynamics. The luminophores decay in parallel: the fastest decay earlier and all the others follow according to a hierarchical order based on the characteristic relaxation time $\tau$.

For the compressed range, $\beta=1.5$, the reduced mass $\mu(t)$ behaves in a markedly different way. As shown in Fig. \ref{fig2}, the compressed $\mu(t)$ presents a convexity which is absent in the stretched case. The curve decreases at the beginning and once it reaches its minimum, just before $t$ approaches the value $1$, it starts to increase almost exponentially. Physically, this trend suggests that there is a serial mechanism: a pumping followed by a damping mechanism. The dynamics is no single any more but it is a dual dynamic. The attention should be focused on the minimum where the system reaches a sort of quasi-equilibrium as a consequence of the reconfiguration or reorganization of the luminophores. Considering a series expansion of the KWW function, it is possible to have a deeper insight. This can explain the reconfiguration and its physical manifestation as a plateau in the time-resolved luminescence only when $\beta$ is in the compressed range.

\section{Physicol-chemical interpretation of the KWW function as a sum of processes}\label{sez3}
The anomalous behavior of the KWW function has been shown to be the result of a mechanism based on a damped harmonic oscillator. However, as hinted in the Introduction, there are two ranges of $\beta$, and therefore there are two KWW functions and two dynamics, so that it is also expected to find two different origins behind the damping. The aim of this Section is to understand how, and especially in which cases, these differences emerge using the series expansion of the KWW function, the so-called Prony series \cite{c41,c42}. The Prony series reproduces quite faithfully the behavior of the KWW function using a finite sum of simple exponential functions: 
\begin{equation}\label{11}
    e^{-\Big(\frac{t}{\tau}\Big)^\beta}\sim\sum_{i=0}^{N}A_ie^{-K_i{t}},
\end{equation}
weighted by the coefficients $A_i$ that should only satisfy the following constraint:
\begin{equation}\label{12}
    \sum_{i=0}^{N}A_i=1.
\end{equation}
{In what follows, we will consider only the zero-th and the first-order terms in the Prony series expansion. The choice to approximate the KWW function as an only two-terms Prony series is due to the fact that usually photoluminescence processes are quite satisfactory described as a bi-exponential model in a first run. And it justifies the use of this tool instead of other more {sophisticated ones as illustrated in} \cite{c51,c52}. Moreover, we need to consider this series expansion as a \textit{lens} on the processes underlying the overall anomalous behavior described by the KWW function. In other words, we want to \textit{zoom} on the system and understand the different dynamical origins of the two KWW functions. Before we proceed to the physical interpretation of the terms, we need to convince the reader that a two-terms Prony series is enough for the forthcoming discussion and therefore the best fit between the KWW function and the Prony series for the parameters involved in \eqref{12} can be summarized as follows.}

The KWW function approximates the stretched exponential function if $K_i$ are real and $A_i$ are real and positive, whereas the KWW function mimics the compressed behavior if $K_i$ are real but $A_i$ are real and of opposite signs. Moreover, the parameters $K_i$ in the stretched case differ by several order of magnitude, as emerged from Tab. 1, and such difference increases as $\beta$ tends to zero. 
\begin{table}[ht]
\centering
\caption{Parameters of the two-terms Prony series that approximates the stretched KWW function for $\beta<1$. The last column is the sum of the squared of the residuals (SSRs) $\Sigma_j(y_j-f(x_j))^2$ between the KWW function and the Prony series.}\label{tab_stretched2}
\begin{tabular}{ccccccc}
\toprule
\multirow{1}*{$\beta$}& \multicolumn{4}{c}{parameters}&\multirow{1}*{SSRs}\\
\midrule
&$A_1$&$A_2$& $K_1$ &$K_2$&&\\
\midrule
0.1&0.906071 &0.0939289& 0.356131 &1.0391~$10^{-5}$ & 10.9237\\
\midrule
0.2&0.891382 &0.108618& 0.71233 &2.1053~$10^{-3}$ & 5.24985~$10^{-1}$\\
\midrule
0.3&0.808055 &0.191945& 1.2683 &3.1232~$10^{-2}$ & 3.25275~$10^{-2}$\\
\midrule
0.4&0.749713 &0.250287& 1.56817 &9.9424~$10^{-2}$ & 4.84242~$10^{-3}$\\
\midrule
0.5&0.694155 &0.305845& 1.73787 &1.9646~$10^{-2}$ & 9.52931~$10^{-4}$\\
\midrule
0.6&0.63105 &0.36895& 1.83565 &3.1490~$10^{-1}$ & 1.99314~$10^{-4}$\\
\midrule
0.7&0.552047 &0.447953& 1.88862 &4.5187~$10^{-1}$ & 3.83391~$10^{-5}$\\
\midrule
0.8&0.445005 &0.554995& 1.90991 &6.0787~$10^{-1}$ & 5.68391~$10^{-6}$\\
\midrule
0.9&0.284997 &0.715003& 1.90434 &7.8697~$10^{-1}$ & 4.29233~$10^{-7}$\\
\bottomrule
\end{tabular}
\end{table}

On the other hand, Tab. 2 shows that the parameters $K_i$ in the compressed case are of the same order of magnitude. 

\begin{table}[h!]
\centering
\caption{Parameters of the two-terms Prony series that approximates the compressed KWW function for $\beta>1$. The last column is the sum of the squared of the residuals (SSRs).}\label{tab_compressed}
\begin{tabular}{ccccccc}
\toprule
\multirow{1}*{$\beta$}& \multicolumn{4}{c}{parameters}&\multirow{1}*{SSRs}\\
\midrule
&$A_1$&$A_2$ &$K_1$ &$K_2$& & \\
\midrule
 1.1&9.27217 &-8.27217&0.79703 &0.773835 & 8.98102$\cdot10^{-5}$\\
\midrule
1.2&13.827 &-12.827&0.758224 &0.73983 & 4.28344$\cdot10^{-4}$\\
\midrule
1.3&18.1737 &-17.1737 &0.741702 &0.726466 & 9.96678$\cdot10^{-4}$\\
\midrule
1.4&19.5284 &-18.5284 &0.734959 &0.719935 & 1.7287$\cdot10^{-3}$\\
\midrule
1.5&21.9199 &-20.9199& 0.731631 &0.717709 & 2.55749$\cdot10^{-3}$\\
\midrule
1.6&22.4522 &-21.4522 &0.730799 &0.716776 & 3.42634$\cdot10^{-3}$\\
\midrule
1.7&23.8176 &-22.8176 &0.730602 &0.717073 & 4.2894$\cdot10^{-3}$\\
\midrule
1.8&24.809 &-23.809 &0.730919 &0.717692 & 5.11091$\cdot10^{-3}$\\
\midrule
1.9&27.0386 &-26.0386 &0.731071 &0.718768 & 5.86421$\cdot10^{-3}$\\
\midrule
2.0&26.5933 &-25.5933 &0.731862 &0.719196 & 6.53116$\cdot10^{-3}$\\
\bottomrule
\end{tabular}
\end{table}
\newpage
However, expanding the KWW function into a Prony series does not allow an immediate understanding of the physical meaning and the role of $A_i$ and $K_i$ in the dynamics of the PL emission. The key to unlock their meaning is to use the physicol-chemical interpretation based on the kinetic equations.\\
Without loss of generality, we will consider only a few decay pathways, as shown by the three-energy level scheme in Fig. \ref{fig3}.

\clearpage
\begin{center}
\includegraphics[scale=0.2]{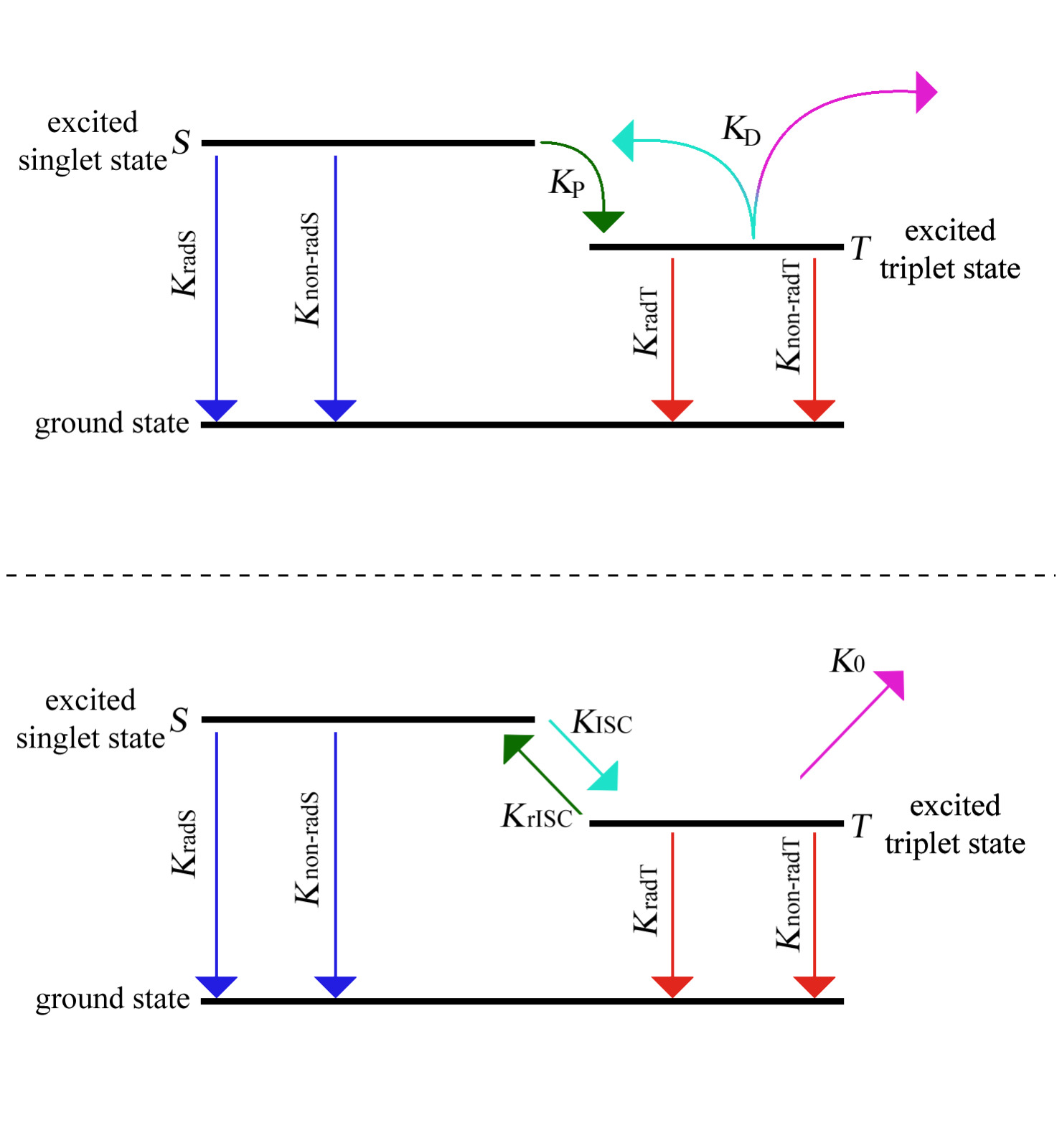}
\captionof{figure}{Energy diagram showing the fundamental decay rates used to model a photoluminescent system. Excited vibrational levels have been omitted for simplicity, as well as the pumping from the ground state. }
    \label{fig3}
\end{center}

The time evolution of the excited luminophores in the singlet and triplet states is given by the following two coupled equations:
\begin{equation}\label{13}
\begin{cases} \dfrac{dn_S(t)}{dt}=-K_S~n_S(t)+ K_D~n_T(t), ~~&n_S(0)=m\\ \dfrac{dn_T(t)}{dt}=K_P~n_S(t)-K_T~n_T(t),~~&n_T(0)=n, \end{cases} 
\end{equation}
where $n_S(t)$ and $n_T(t)$ denote respectively the fraction of the luminophores in the singlet or triplet excited state at the time $t$ whereas $m$ and $n$ represent the fraction of the luminophores at the initial time $t=0$, respectively. The constraint to be fulfilled is $n_S(0)+n_T(0)=1$.

Here, the triplet state is a special trap where the excited electron can be caught after inverting its spin. $K_{\mbox{rad}}$ and $K_{\mbox{non-rad}}$ denote the radiative and non-radiative rates from $S$ (or $T$) to the ground state $S_0$, respectively. $K_{\mbox{radS}}=\frac{1}{\tau_F}$ determines the rate of the fluorescence whereas $K_{\mbox{radT}}$ is the rate of the phosphorescence. Thus, $K_S=K_{\mbox{radS}}+K_{\mbox{non-radS}}$ and $K_T= K_{\mbox{radT}}+K_{\mbox{non-radT}}$.  $K_P$ is the total trapping rate, and in case of a photoexcited system it coincides with the intersystem crossing rate, $K_P=K_{ISC}$, that is the rate at which the triplet state is populated by the singlet excited state. $K_D$ is the total de-trapping rate and it is defined as the difference between the reverse intersystem crossing rate, i.e. the re-population of the singlet excited state from the triplet one $K_{\mbox{rISC}}$, and the triplet-trap rate $K_O$, which is the rate at which the excited triplet state is impoverished via interaction with an external trap due to the presence of more active species, as oxygen for instance. Finally, $K_D = K_{\mbox{rISC}} - K_O$. 

The system \eqref{13} can be solved for both excited states. However, since we are interested in the fluorescence intensity emission that is proportional to the number of luminophores in $S$, we will consider only the solution that describes the time evolution of the number of the luminophores in that excited state. The solution is a bi-exponential function of the form:
\begin{equation}\label{14}
    \begin{split}
        n_S(t)&=a_1~e^{-k_1~t}+a_2~e^{-k_2~t},
    \end{split}
\end{equation}
where the total decay rates $k_i$ and the coefficients $a_i$ are given by:
\begin{itemize}
    \item[] $k_1=\dfrac{K_S+K_T+\sqrt{(K_S-K_T)^2+ 4K_DK_P}}{2}$,
    \item[] $k_2=\dfrac{K_S+K_T-\sqrt{(K_S+K_T)^2+4 K_DK_P}}{2}$,
    \item[] $a_1=\dfrac{m(K_S-K_T)+m\sqrt{(K_S-K_T)^2+ 4K_DK_P}-2K_D n}{2\sqrt{(K_S-K_T)^2+4 K_DK_P}}$,
    \item[] $a_2=\dfrac{-m(K_S-K_T)+m\sqrt{(K_S-K_T)^2+ 4K_DK_P}-2K_D n}{2\sqrt{(K_S-K_T)^2+4 K_DK_P}}$.
\end{itemize}
According to the above definition of the $a_i$ coefficients, the bi-exponential function \eqref{14} results to be normalized, i.e. $a_1+a_2=1$, for any set of values of $K_D$, $K_P$, $K_S$ and $K_T$. Therefore, it can be considered as a two terms Prony series expansion of the KWW function \eqref{11}:
$$a_1=A_1,a_2=A_2$$ and $$k_1=K_1, k_2=K_2.$$ In other words, using the physicol-chemical interpretation based on only few decay routes we obtain a definition of the PL emission that represents the truncation of the series expansion of the KWW function. We can thus identify \eqref{1} with \eqref{14}.

However, to be physically consistent, the decay rates should respect the constraints listed below: 
\begin{subequations}\label{15}
\begin{align}
&K_S>K_T\geq0,\label{a-cond} \\
&K_{\mbox{ISC}}>>K_S\label{b-cond}\\
&K_{\mbox{rISC}}>>K_T\label{c-cond}\\
&K_{\mbox{ISC}}>>K_{\mbox{rISC}}\label{d-cond}\\
&K_{\mbox{non-rad T}}>>K_{\mbox{rad T}} \label{e-cond}.
\end{align}
\end{subequations}
The decay rate $K_S$ is greater than the triplet decay rate $K_T$ (15a) due to the fact that the fluorescent emission is a faster light-emitting process than phosphorescence. The constraint on the intersystem crossing (15b) (or the reverse intersystem crossing (15c)) prevails on the de-activation of the excited states. Thus, explains the excitons crossover between the $S$ and $T$ states. However, as established by the Hund’s rule \cite{c43}, the triplet state $T$ is always energetically lower than the singlet excited state $S$. Therefore the intersystem crossing is more favorable than the reverse process (15d). To guarantee the reverse intersystem crossing, two conditions should be fulfilled: the energy split $\Delta E_{ST}$ between $S$ and $T$ should be small and an endothermic process, as the thermal activation, should be involved to overcome the energy gap. Under these requirements, the $S$ state can be re-populated with a delay compared to the time needed to populate the $S$ state directly via photoexcitation. The last constraint (15e) imposes a direction to the competition between the non-radiative and radiative decay rates. This last inequality explains why often the phosphorescence is not observed at room temperature. However, it is important to underline that the reverse intersystem crossing decay rate $K_{\mbox{rISC}}$ is strongly dependent from the temperature according to the Arrhenius law \cite{c44,c45,c46}:
\begin{equation}\label{16}
K_{\mbox{rISC}}=Ce^{-\frac{\Delta E_{ST}}{k_BT}},
\end{equation}where $C$ is a factor related to the spin-orbit coupling, $\Delta E_{ST}$ the singlet-triplet energy gap, $k_B=8.617333262\cdot10^{-5}$ev/K the Boltzmann constant, and $T$ the temperature measured in degrees Kelvin. Therefore, the de-trapping rate $K_D$, defined as a difference between $K_{\mbox{rISC}}$ and $K_O$, can be both positive or negative. This is the key factor behind the presence of the stretched and the compressed KWW function. The sign of $K_D$ influences the sign of the coefficients $A_1$ and $A_2$ in \eqref{14} and determines the stretched or compressed nature of the KWW function.  The presence of negative coefficients $A_i$ indicates that the system is almost entirely decayed. So the compressed behavior is not physically admissible in the first part of the time-resolved PL. In other words, the compressed exponential behavior is expected only in the last part of the time evolution, when the increasing number of the quenchers starts to influence greatly the decay by accelerating the degradation of the photoluminescent system.

\section{An ideal experiment. The KWW function as a reaction kinetic process.}\label{sez4}
At this point, we give a validation to the results and considerations illustrated in the previous Sections. We show how they can naturally emerge in a fairly straightforward way considering the evolution of an ideal photo-excited system. 

After switching off the initial photoexcitation, all the luminophores are in the singlet excited state and they start to decay and relax to the ground state according to the bi-exponential model \eqref{14}. Mathematically, the initial conditions for system \eqref{13} are $n_S(0)=m$ and $n_T(0)=0$, where $m$ is a finite number, less or equal to $n_0$, that represent a fraction of the luminophores.

As shown in Fig. \ref{fig3}, the luminophores in the excited singlet state $S$ decay to the ground state or to the triplet excited state $T$, which is initially empty and therefore it begins to be populated via intersystem crossing. Whatever the path chosen by the excitons, the singlet excited state is depleted. In this situation, both $A_1$ and $A_2$ assume only positive values and respect the normalization condition \eqref{12}. This result perfectly describes the physical situation, that is, the singlet excited state is decaying to the ground state and via intersystem crossing, whereas the reverse intersystem crossing and the other radiative and non-radiative pathways are present but still asleep. Combining the positiveness of both coefficients and the aforementioned decay routes, the dynamics is dominated by a damping mechanism and the bi-exponential model follows a stretched KWW behavior confirming what has been stated at the end of Section 3, that is, the stretched exponential function dominates the early stages of the dynamics.

At this point, both excited states are populated and then, other paths illustrated in Fig. \ref{fig3} are available. The conditions on the parameters involved in the definition of $A_1$ and $A_2$ to be satisfied are:
\begin{subequations}\label{17}
\begin{align}
K_S&>K_T\geq0, \\
K_D&=K_{\mbox{rISC}}-K_O\leq0\\
K_P&>0\\
K_P&\leq\dfrac{(K_S-K_T)^2}{4K_D}.
\end{align}
\end{subequations}
Among the paths that are now active, there are the reverse intersystem crossing $K_{\mbox{rISC}}$ and the triplet-trapping rate $K_{\mbox{rISC}}=K_O$. As known in Section 3, $K_D=K_{\mbox{rISC}}-K_O$ can be positive or negative. In case $K_D$ is positive, the pumping mechanism is acting on the singlet excited state. In other words, the reverse intersystem crossing dominates the dynamics and push the luminophores confined in the triplet state to decay to the singlet excited state and then, relax producing a delayed luminescence. This delay in the luminescence can be appreciated considering the difference in the order of magnitude found in the Prony expansion of the stretched KWW function (see Tab. 1)

When $K_{\mbox{rISC}}=K_O$ (or i.e. $K_D=0$), the bi-exponential reduces to a simple mono-exponential function. This implies the presence of a plateau in the double-log plot highlighting a re-configuration of the luminophores, that is, a change of their spin. At the same time, it suggests that the pumping action of the $K_{\mbox{rISC}}$ is inhibited by the increasing influence of $K_O$. 

As the temperature goes down, $K_{\mbox{rISC}}$ tends to zero whereas the triplet-trapping rate grows up due to the increasing of the concentration of the quenchers $K_D=K_{\mbox{rISC}}-K_O<0$, and then the damping mechanism dominates the dynamics. The triplet and the singlet state are in quasi-equilibrium, or in other words, the production rate of the triplet state is zero. Then, the presence of the pumping mechanism can be definitely neglected if the non-radiative triplet-trap decay rate $K_O$ is more competitive than the reverse intersystem crossing rate $K_{\mbox{rISC}}$. Therefore, the concentration of the quenchers starts to be influential on the dynamics of the system. The pumping mechanism has just stopped its action, $K_D<0$, and the decay of the number of luminophores is unstoppable. Under these conditions, the system \eqref{13} can be reduced to:
\begin{equation}\label{18}
\begin{cases} \dfrac{dn_S(t)}{dt}=-K_S~n_S(t)- K_O~n_T(t), ~~&n_S(0)=m\\ \dfrac{dn_T(t)}{dt}=0,~~&n_T(0)=n. \end{cases} 
\end{equation}
which has the solution:
\begin{equation}\label{19}
    n_S(t)=\Big(m+\frac{K_O}{K_S}n\Big)e^{-K_St}-\frac{K_O}{K_S}n.
\end{equation}

As expected, it can be rewritten in terms of a two terms-Prony series where the parameters and coefficients are given by
\begin{itemize}
    \item[] $K_1=K_S$ and~$A_1=m+\dfrac{K_O}{K_S}n$,
    \item[] $K_2=0$ ~~~and~$A_2=-\dfrac{K_O}{K_S}n$,
\end{itemize}
which fulfill the constraint of \eqref{12} imposed on the Prony series coefficients.

The solution in (19) can be divided into two parts, a 0th order kinetics and a 1st order kinetics. The 0th order kinetics is a constant rate that is dependent on the concentration of luminophores in the triplet state, and it is independent on the concentration of the luminophores in the singlet excited state. On the contrary, the 1st order kinetics is directly proportional to the concentration of luminophores in the singlet state. This solution is consistent with the one obtained in \cite{c2}, where the photoluminescent system is modelled as a combination of 1st order and 0th order kinetics.

\section{Requirements for modelling relaxation}\label{sez5}
The term \textit{component} is introduced to indicate the ordinate spatial aggregation inside the class of the luminophores that shows the same PL degradation decay (i.e. the same value of $\beta$) and consequently, also the same dynamics. In the results obtained up to now, there are two classes of luminophores: one described by the stretched KWW function and the other one by the compressed KWW function. The introduction of these components inside the classes do not imply a long-range spacial grouping, but it describes ensembles whose elements are locally distributed on the entire system. In other words, a component is a nonlocal ensamble of luminophores that have the same life-cycle. Despite their distance, the emitting centers that belong to the same component have the same time-resolved photoluminescence. The aggregation in components can be interpreted in the light of the mathematical properties of monotonicity and complete monotonicity. These properties can be used as a requirement in modeling the different dynamics in PL decay since they embed all the results and observations that allows to distinguish the two KWW functions. The following definition will be used from now on \cite{c47}. 

A function is called monotonically increasing (or decreasing) in a domain, if for all $x$ and $y$ such that $x\leq y$, $f(x)\leq f(y)$ ($f(x)\geq f(y)$). It is called complete monotone on a domain $I$, if it has derivatives of all orders that satisfies the condition: 
\begin{equation}\label{20}
    (-1)^rf^{(r)}(x)\geq 0, ~r\in\mathbb{N}_0, ~x\in I,
\end{equation}
where $f^{(r)}$ represents the r-th derivative with respect to $x$ of the function $f(x)$. Therefore, a complete monotone function is non-negative as it is immediate to conclude considering the case $r=0$ in \eqref{20}. 

By applying the above definition, it is evident that the compressed KWW function satisfies the requirements needed to be a monotonic function, whereas the stretched KWW function is completely monotone \cite{kww04}. In this last case, the proof can be obtained either via Laplace transform and Bernstein theorem \cite{c48,c49} {or directly by considering the result obtained by Pollard \cite{c50}.} {A last but not least possibility is to apply what we call “domino-effect” algorithm (see Fig. 4 for a graphical sketch of how the mathematical definitions, properties and theorems allows to prove a complete monotonicity for a function under specific initial conditions).}
\\
The "domino-effect" algorithm guarantees the complete monotonicity of the $f(x)$, if the $f(x)$ is a positive, strictly decreasing and convex function such that it belongs to the class $C^k(I,\mathbb{R})$, where $I=]0,\infty[$ is an open subset of the real number $\mathbb{R}$. \\
At this point, the complete monotonicity of the stretched KWW function arises by chain conditions between higher orders derivatives as a consequence of the application of basic theorems on concave and convex functions. 

\begin{center}
\includegraphics[scale=0.45]{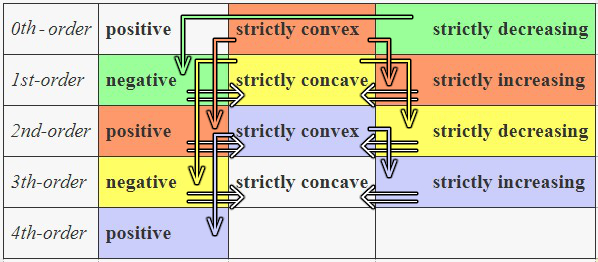}\label{domino}
\captionof{figure}{The \textit{domino} effect algorithm.}
\end{center}

In conclusion, the range of the parameter $\beta$ plays the role of a watershed: the KWW function can be monotone or completely monotone and it is possible to associate respectively the single and dual dynamics observed in Sec. 2. 

The stretched KWW function describes the class of completely monotone luminophores revealing the single nature of the PL emission decay: only the damping mechanism governs the overall dynamics. This aspect is important because it guarantees that the luminophores in the class follow parallel decay routes, or in other words the damping mechanism dominates on the time-evolution PL decay, resulting into parallel daughters’ pathways. Here, the luminophores follow pathways that exist from the beginning: the fluorescent decay, the non-radiative excited singlet decay and the intersystem crossing. Other possibilities are still not active. The luminophores do not interact with other species, as molecular dioxygen or more active molecules in the environment when the dynamics is ruled by a complete monotone relaxation function.  On the other hand, the class of monotone luminophores describes a serial or strictly sequential decay process. Here, the dynamics has a dual nature: there is a pumping mechanism followed by a damping mechanism, as obtained analyzing the reduced mass $\mu(t)$ in the compressed range. The serial nature of the monotonic decay can be physically appreciated by the presence of a plateau in the luminescent emission. This plateau testifies the re-organization experienced by simple monotone components, and it is a remarkable difference from the components belonging to the complete monotone luminophores class in which the plateau is absent. The monotone dynamics described by the compressed KWW function is not a self-consistent process, as the stretched KWW decay, but it has to follow the single dynamics since the pathways involved are activated only after the damping mechanism ruled by the stretched KWW function. Here, the number of the quenchers or their concentration grows up in the sample, so their influence becomes more and more important until the decay is unstoppable and increasingly faster with the consequent degradation of the sample itself. The destructive effect of the increasing concentration of quenchers has been observed considering the Prony series expansion of the KWW function where the main differences characterizing the two classes of luminophores clearly emerges. In the Prony series of the stretched KWW function all the terms are positive, whereas in the Prony series that approximates the compressed KWW function there are also negative terms. The series expansion of the stretched KWW function is expected to be a sum of only positive terms since this approximation plays the role of the discretization of the Laplace integral \cite{das}:
\begin{equation}\label{21}
    I(t)=\int_0^\infty g(k)e^{-kt}dt.
\end{equation}
In the case of a complete monotone PL intensity $I(t)$, the integrand function $g(k)$ is a probability density distribution \cite{c50,c51,c52,c53}. On the other hand, the presence of the negative terms mathematically is due to the fact that there are no conditions on its positiveness, as instead required in case of complete monotonicity. In some cases, the integrand function in \eqref{21} can be negative. From a physical point of view, the negative terms are related to the quenching of the luminophores by the molecular dioxygen present during the PL measurements (in fact, they depend on $K_O$). These very reactive species can impoverish the triplet state de-trapping the excited electron, thus favoring the de-activation of the state and preventing the fluorescent emission.

To sum up, the complete monotonicity provides significant information on the dynamics of the relaxation processes ruled by the KWW function. The stretched (complete monotone) KWW function is characterized by a single dynamics which is further linked to a higher order organization of the components that has the physically relevant consequence in the optimization of the PL intensity. As it will be shown in the forthcoming Sections, our interpretation is revealed to be correct. In fact, when a heat treatment, called thermal annealing, has been applied to the whole sample  altering the microstructure of a material in order to change its properties, it improves not only the duration but also the PL intensity itself. The main consequence is the re-organization of the luminophores in a different grouping of components. The new components are described by complete monotone functions. The complete monotonicity is therefore an asset mathematical tool that clearly indicates the presence of an higher-order organization in the sample: the luminophores are better protected from the quenching of the external agents and it is perfectly confirmed by the results obtained via Prony series. On the other hand, the simple monotonic requirement satisfied by the compressed KWW function indicates that the system is becoming more and more heterogeneous (indeed, it is interacting with the environment) and it is forced to degrade. Here the luminophores are not so highly organized as in the other case and it is reflected by the dual nature of its dynamics that need the presence of a serial mechanism: a pumping followed by a damping. In other words, the system is going to fix the disorder favoring the fluorescent emission. In particular, the presence of a plateau in the normalized PL emission is a hint that the system is modeled by a compressed KWW function.

\section{An application: the Alq3 molecule}\label{sez6}
This Section is devoted to the application of the previous results to a real physical photoluminescent system as the Tris(8-hydroxyquinoline)aluminum (Alq3). The choice of this organic molecule is based on two main reasons: the widely-used applications of this material in organic light-emitting diodes (OLEDs) \cite{c27,c28,c29,c31}, a spreading technology in our daily life, and the large amount of data available on its optical properties \cite{c1,c2,c3,c4}.
For instance, the PL emission of thin films of Alq3, annealed at different temperatures and at different wet and dry atmospheres, has been throughout studied as a function of thickness, temperature, excitation wavelength, and decay time in various atmospheric environments. In particular, especially important for the present work, the intensity of PL for very long decays in Air has been measured in detail for twelve samples \cite{c1,c2,c3,c4}. Without loss of generality, we have focused our attention on two of them, the reference sample alq63-3 that has not been annealed, and sample alq65-1 that has been annealed in dry oxygen at 180$^\circ$C.\\ The intensity of PL of these samples is described by the following equation:
\begin{equation}\label{22}
    I(t)=\sum_{i=0}^{4}I_ie^{-\Big(\frac{t}{\tau_i}\Big)^{\beta_i}},
\end{equation}
where the total intensity $I(t)$ consists of four components characterized by three parameters $I_i$, $\beta_i$ and $\tau_i$ ($i=1,...,4$), which represent the amplitude, the KWW parameter and the time constant of the i-th component, respectively. In order to satisfy the normalization condition to the overall process, \eqref{22} should respect the following constraint imposed on the sum of the amplitudes
\begin{equation}\label{23}
    \sum_{i=1}^{4}I_i=1.
\end{equation}
Here, the amplitudes of the components depend on the physical and chemical conditions during the thermal evaporation and annealing processes of the films, whereas the characteristic time constants depend on the environmental conditions during the long degradation in Air.

When $\beta=1$, all the components are pure exponential functions. In order to appreciate how significant are the improvements in the fit quality carried out with the $\beta$ parameter, a comparison between the two best fits resulting from \eqref{22} with $\beta=1$ and $\beta\neq1$ has been performed in the following.

Figures \ref{fig4} and \ref{fig5} illustrate the normalized intensity emission of samples alq63-3 and alq65-1 as a function of decay time in Air.\\ A best fit calculation with the Four Component Model (FCM), produces the solid and dashed curves, which are the four components, with the parameters reported in Table 3.
\begin{center}
\includegraphics[scale=0.5]{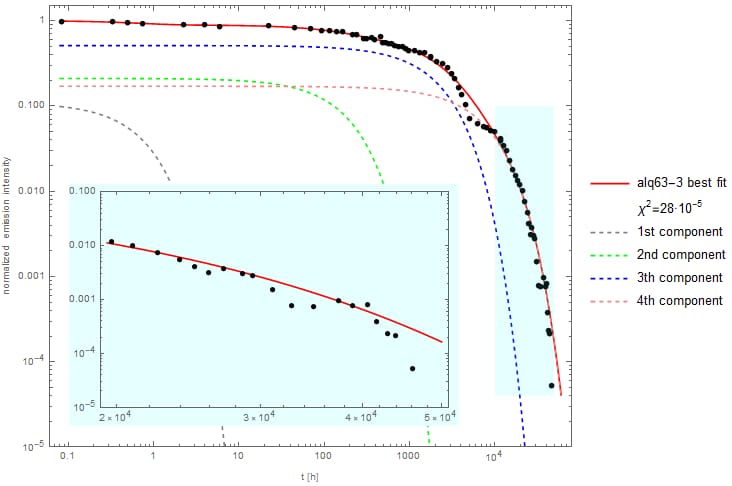}
\captionof{figure}{PL intensity of the non-annealed sample alq63-3 decaying in Air (full black circles), best fit by using \eqref{22} with $\beta=1$ (solid red line) and its four components (dashed lines) according to parameters presented in Table 3. Reproduced from \cite{c2}. The light blue insert reports the same graph above $20,000$h.}
    \label{fig4}
\end{center}
\begin{center}
\includegraphics[scale=0.45]{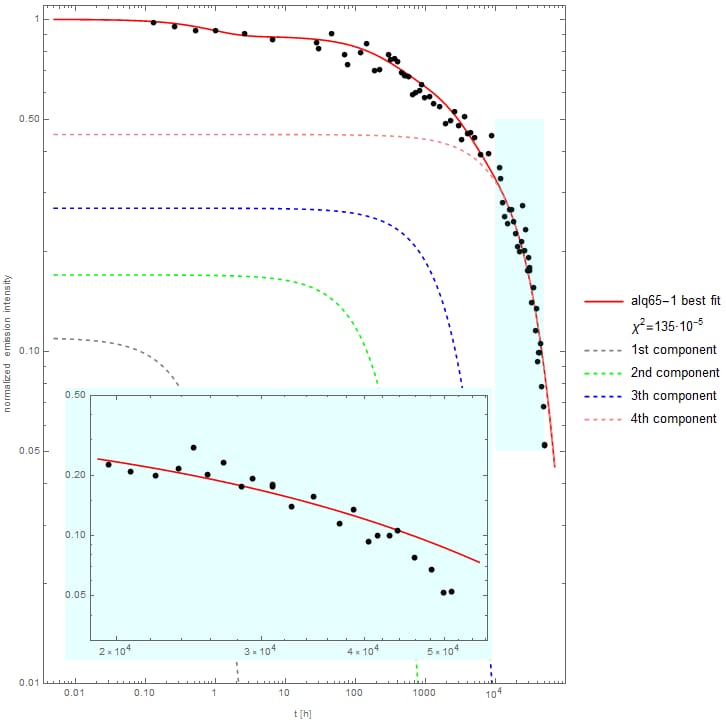}
\captionof{figure}{PL intensity of sample alq65-1 annealed in dry oxygen at $180^\circ$C decaying in Air (full black circles), best fit by using \eqref{22} with $\beta=1$ (solid red line) and its four components (dashed lines) according to parameters presented in Table 3. Reproduced from \cite{c2}. The light blue insert reports the same graph above $20,000$h.}
    \label{fig5}
\end{center}
\begin{table}[ht]
\centering
\caption{Amplitudes and time constants of the four components for the reference not-annealed sample alq63-3 and the sample alq65-1, which has been annealed in dry oxygen at 180$^\circ$C, retrieved by fitting the experimental data with \eqref{22} and $\beta=1$.}\label{tab_final}
\begin{tabular}{cccccccccc}
\toprule
\multicolumn{1}{c}{Sample}&\multicolumn{4}{c}{Amplitude (\%)}&\multicolumn{4}{c}{Time constant(s$^{-1}$)}&\multicolumn{1}{c}{$\chi^2$($10^{-5}$)}\\
\midrule
 &$I_1$ & $I_2$ & $I_3$ & $I_4$ & $\tau_1$ & $\tau_2$ & $\tau_3$ & $\tau_4$  \\
\midrule
alq63-3&11 & 21 & 51 & 17 & 0.73 & 174 & 2079 & 7200 & 28\\
\midrule
alq65-1&11 & 17 &27 & 45 & 0.90 & 276 & 2700& 30,398 & 135\\
\bottomrule
\end{tabular}
\end{table}
Overall, the fit is satisfactory, especially considering that the decay time spans five decades. However, the FCM does not fit the negative bump in the reference sample alq63-3 around $5000$h in Fig. \ref{fig4}, and the fast-experimental decreasing glimpsed after $30,000$h in Figs. \ref{fig4} and \ref{fig5}, and observed much better in their inserts above $20,000$h.
The previous discrepancies between theory and experiments are overcome by using \eqref{22} with $\beta$ as a free parameter, and Figures \ref{fig6} and \ref{fig7} show the best fit calculation, which produces the solid and dashed curves, which are the four components with the new parameters reported in Table 4.
\begin{center}
\includegraphics[width=\linewidth]{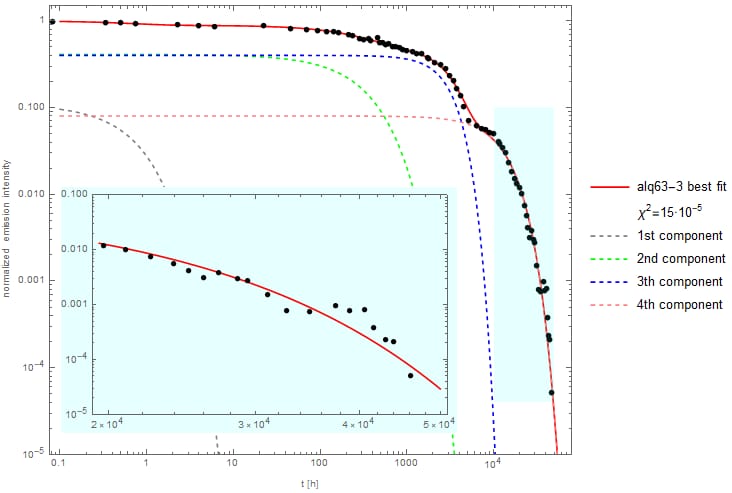}
\captionof{figure}{PL intensity of the non-annealed sample alq63-3 decaying in Air (full black circles), best fit by using Eq. \eqref{22} (solid red line) with $\beta\neq1$ and its four components (dashed lines) according to parameters presented in Table 4. Reproduced from \cite{c2}. The light blue insert reports the same graph above $20,000$h.}
    \label{fig6}
\end{center}
\begin{center}
\includegraphics[scale=0.45]{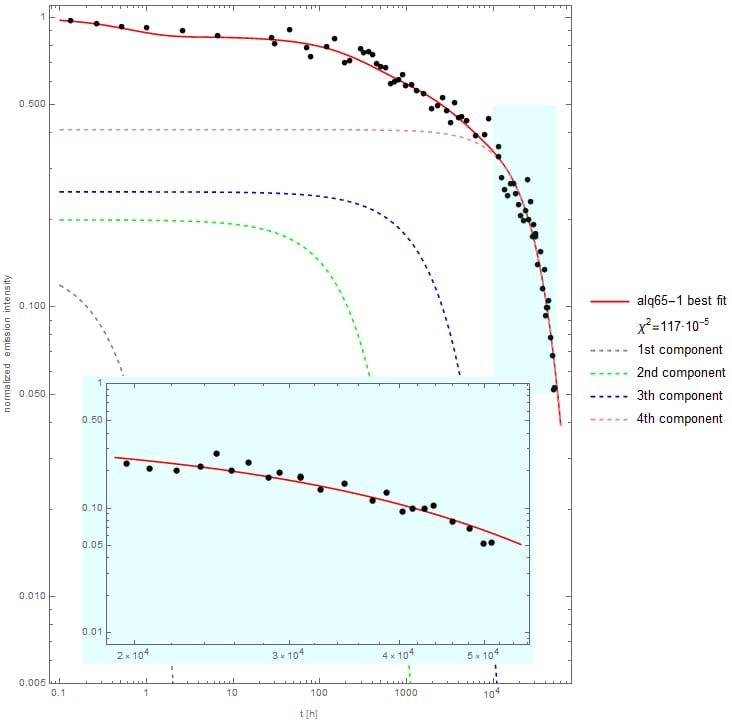}
\captionof{figure}{PL intensity of sample alq65-1 annealed in dry oxygen at $180^\circ$C decaying in Air (full black circles), best fit by using \eqref{22} (solid red line) ) with $\beta\neq1$  and its four components (dashed lines) according to parameters presented in Table 4. Reproduced form \cite{c2}. The light blue insert reports the same graph above $20,000$h.}
    \label{fig7}
\end{center}

\begin{table}[ht]
\centering
\caption{Amplitudes, time constants, and KWW parameter $\beta$ of the four components for the reference not-annealed sample alq63-3 and the sample alq65-1, which has been annealed in dry oxygen at 180$^\circ$C, retrieved by fitting the experimental data with \eqref{22}.}\label{tab_final}
\begin{tabular}{cccccccccccccc}
\toprule
\multicolumn{1}{c}{Sample}&\multicolumn{4}{c}{Amplitude ($\%$)}&\multicolumn{4}{c}{Time constant(s$^{-1}$)}&\multicolumn{4}{c}{$\beta$}&\multicolumn{1}{c}{$\chi^2$($10^{-5}$)}\\
\midrule
 &$I_1$ & $I_2$ & $I_3$ & $I_4$ & $\tau_1$ & $\tau_2$ & $\tau_3$ & $\tau_4$ & $\beta_1$ & $\beta_2$ & $\beta_3$ & $\beta_4$ & \\
\midrule
alq63-3&11 & 41 & 40 & 8 & 0.72 & 332 & 3195 & 13381 & 1 & 1 & 2 & 1.57 & 15\\
\midrule
alq65-1&14 & 20 &25 & 41 & 0.61 & 300 & 2819 & 30,398 & 1 & 1 & 1 & 1.39 & 117\\
\bottomrule
\end{tabular}
\end{table}
Certainly, a single glance is enough for understanding the improvement in the quality of the fit in both samples. Indeed, the red solid line is able to better interpolate the experimental data as it reproduces the bump and the normalized PL intensity for times greater than 30,000 h, which is better observed in the inserts. A quantitative evidence is furnished by the chi-squared values $\chi^2$, which are reported in Figs. \ref{fig4}-\ref{fig7} and Tables 3 and 4. For both samples considered, the values are smaller in case $\beta>1$, which means that four KWW functions model much better the four PL components. Consequently, a satisfactory overlap between experimental data and final best fit is achieved.\\
A last but not least comment concerns the end tails of the normalized PL emission, which the inserts of Figs. \ref{fig4}-\ref{fig7} display with magnification. In both Alq3 samples considered, the first three components are destroyed and the PL intensity emission is governed by the fourth component, which is the only one still surviving after $10,000$h. In Figs \ref{fig8} and \ref{fig9}, besides the previous FCM fit with $\beta\neq1$ (solid red line), the experimental data have been fitted with the two Prony series of \eqref{14} (dashed purple line) and (19) (dashed orange line). 
\begin{center}
\includegraphics[scale=0.4]{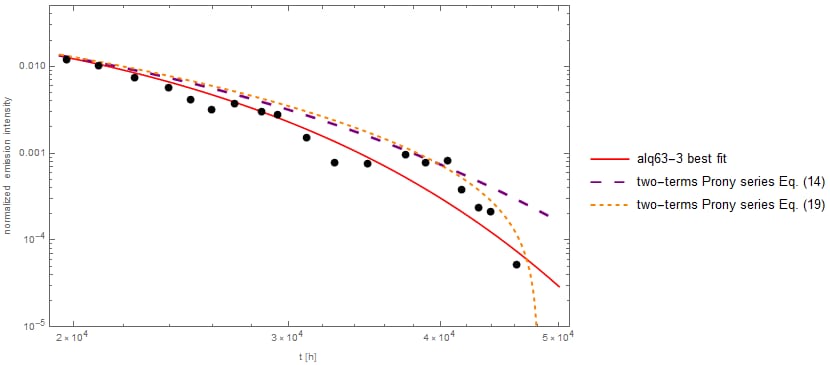}
\captionof{figure}{PL intensity of sample alq63-3 decaying in Air after $10,000$h (full black circles), best fit (solid red line) by using \eqref{22} and parameters presented in Table 4 and the Prony series approximation of the fourth component by using \eqref{14} (dashed purple line) and Eq. (19) (dashed orange line).}
    \label{fig8}
\end{center}

\begin{center}
\includegraphics[scale=0.4]{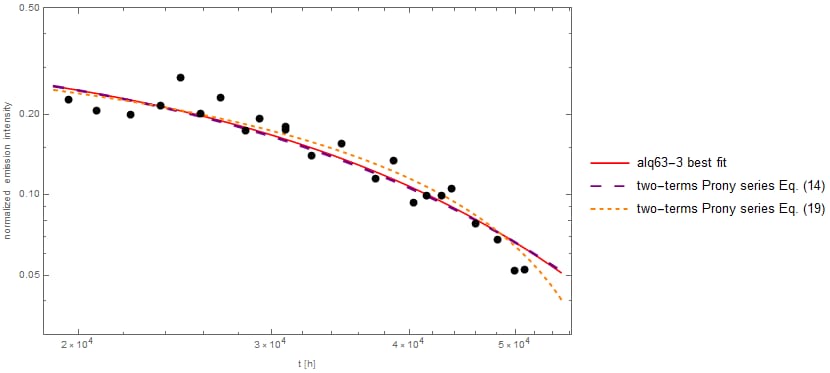}
\captionof{figure}{PL intensity of sample alq65-1 decaying in Air after $10,000$h (full black circles), best fit (solid red line) by using Eq. \eqref{22} and parameters collected in Table 4 and the Prony series approximation of the fourth component by using Eq. \eqref{14} (dashed purple line) and Eq. (19) (dashed orange line). }
    \label{fig9}
\end{center}
It is immediately observed that the two-terms Prony series is a reliable approximation tool for the KWW function even for a physical real system, so confirming the mathematical results obtained exploiting the minimum square method in Section 3. In particular (19), which assumes the total absence of any pumping mechanism, seems to be in better agreement with the experimental data. The parameters of the Prony series (19) obtained from the best fit are reported in Table 5. 
\begin{table}[ht]\label{table_prony_baldacchini}
\centering
\caption{Amplitude and time constants of the fourth component $I_4$ for the reference not-annealed sample alq63-3 and alq65-1, which has been annealed in dry oxygen at 180$^\circ$C, retrieved by fitting the experimental data with Eq. (19).}
\begin{tabular}{cccc}
\toprule
\multicolumn{4}{c}{ Prony series approximation parameters}\\
\midrule
samples&amplitude($\%$)&$K_S${[$10^{-3}$s$^{-1}$]}&$K_D${[$10^{-7}$s$^{-1}$]}\\
\midrule
alq63-3&15&0.12& 0.62~$10^{-7}$ \\
\midrule
alq65-1&40&0.013&67.0~$10^{-7}$\\
\bottomrule
\end{tabular}
\end{table}

It is to be noted that the best fit (solid red line) and the Prony approximation (19) (dashed orange line) tend to overlap. Indeed, the discrepancies between the two lines is no larger than 0.0005! The fitting values in Table 5 are in accordance with the ones collected in Table 3 in reference \cite{c2}, where the singlet decay rate $K_S$ is equal to $\frac{1}{\tau_4}$ and $K_O$ is equal to the parameter $\alpha_4$ introduced in \cite{c2}. Framing these values in the physical context of the excited states sketched in Fig. \ref{fig3}, it is possible to interpret the $\alpha$ parameter introduced in \cite{c2} as the triplet-trap decay rate due to more reactive species as the molecular oxygen that inhibits the luminophores. Finally, the dashed purple lines in Figs. \ref{fig8} and \ref{fig9} represent the solution of the two-terms Prony series \eqref{14} and it is drawn only to highlight the quality of the underlying assumptions in (19). All these facts confirm once again the correctness of the physical framework based on the pumping and damping mechanisms.

{Let us conclude this Section, by pointing out a couple of observations even on the thermal annealing and its consequences. The thermal annealing is a heat treatment process which alters the microstructure of a material in order to change its mechanical and/or electrical properties. The thermal annealing of the emissive layer in an OLED improves incredibly not only the efficiency of the devices but also the PL emission and the lifetime contrasting the degradation for very long time, more than 10,000h.} {To get a more concrete sense we need to compare the intensity decay of the two samples, before and after annealing}. It is clearly evident that the thermal treatment significantly improved the PL emission. Indeed, the normalized emission intensity in the sample alq63-3 after $50,000$h of decaying in Air is about $3\cdot10^{-5}$, whereas is $0.02$ in the annealed sample alq65-1, almost three orders of magnitude bigger! More detailed information on the annealing processes as a mean to improve the radiative yields of films of organometallic compounds is contained elsewhere \cite{c1,c2}.  Another consequence of the thermal annealing is the re-arrangement of the components, which is better deduced by observing the behavior of the components (dashed lines) illustrated in Figs. \ref{fig4} and \ref{fig5} (or Figs. \ref{fig6} and \ref{fig7}) and Table 3 (or Table 4). In sample alq63-3, the order among the components has not been preserved during the photoluminescent decay. Up to $100$h, the order of the intensity arranged from the lowest to the highest is $I_1$, $I_4$, $I_2$ and finally $I_3$, while between $100$h and $1500$h, $I_2$ and $I_4$ swap. Only after $6000$h the order is restored, as remarked in Fig. \ref{fig4} (or in Fig. \ref{fig6}). Instead, in the annealed sample alq65-1, all four components are perfectly organized from the very beginning till the end of the PL decay, i.e. from a fraction of a second to $50,000$h, as remarked by the dashed lines in Figs. \ref{fig5} and \ref{fig7}.  The thermal annealing offers an improvement in the PL emission and minimizes the effect of the disorder re-configuring the orientation of the molecules. The luminophores are better protected from internal and external agents, and therefore they can emit for a longer time.  The luminophores are trapped in the triplet state and their interactions are almost inhibited so that they cannot be easily quenched or decayed to the ground state improving the PL emission.\\
All the above lead us to conclude that the final stages of the system evolution should be described by a compressed KWW function.

\section{Single and dual dynamics in a real physical system}\label{sez7}
Let us apply the approach delineated in Section 2 to investigate the hidden dynamics of the PL emission in terms of pumping and damping mechanism.  As shown in Section 2, these mechanisms are governed by the time-dependent reduced mass $\mu(t)$. Each component has its own reduced mass and their superposition determines the dynamics of the PL intensity emission, and provides an explanation for the plateau and the bumps observed in the plots collected in Section 6. Further confirmation on its correctness arises comparing the results with the dynamical approach via the Prony series expansion based on physico-chemical reactions as shown in Sect. 6.

{Since the thermal annealing boosts the PL intensity and extends its duration, the luminophores must be re-organized. This thermal treatment, essential to be competitive in the production of cutting-edge technologies, flattened the anomalous behaviors as the abrupt damping and the plateau in the luminescence.\\
This leaves the sample alq63-3 as the key reference sample where both these experimental features are more easily visible than in the annealed sample alq65-1. However, as it will be shown in the forthcoming discussion, the same conclusions and observations are applicable to the annealed sample. In fact, once we extricate the complex dynamics and we understand deeply the physical meaning, the results here obtained have an asset value in modeling the time-resolved behavior of photoluminescent material.} Figure \ref{fig10} illustrates the reduced mass of all the four components for the reference sample alq63-3. 
\begin{center}
\includegraphics[width=\linewidth]{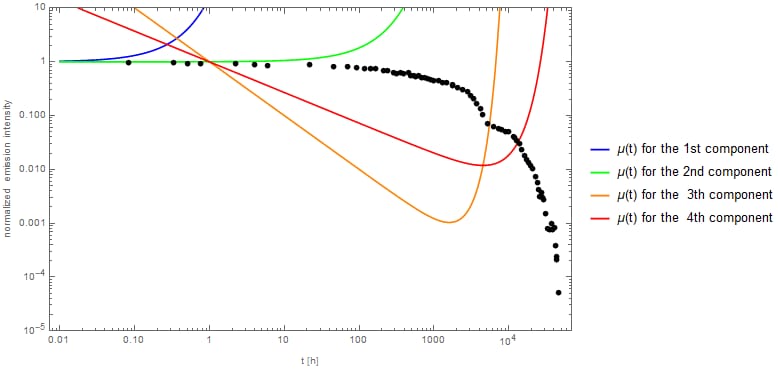}
\captionof{figure}{PL intensity of the sample alq63-3 decaying in Air (full black circles) and the time-dependent reduced mass at different relaxation times corresponding to each component: $\tau=0.72$h solid blue line, $\tau=332$h solid green line, $\tau=3195$h solid orange line and finally $\tau=13,381$h solid red line.}
    \label{fig10}
\end{center}
According to the behavior of the reduced masses (9a), the first two components obey to the single dynamics and contribute only with a damping mechanism whereas the last two components clearly show a dual dynamics, i.e. there are both a pumping and a damping mechanism. This is evident due to the presence of the convexity that is a hallmark of the dual dynamics. The plateau in the luminescence indicates the transition from the pumping to the damping mechanism and in that time window the intensity is almost constant or better it is logarithmically constant. It means that the luminescence decays but its behavior is modeled as a pure exponential function and therefore it is slower then the compressed one. In order to explain the plateau around $5000$h, we should select a proper time range and observe the behavior of each reduced mass $\mu(t)$ inside the range. Between $4500$h and $13,385$h, only the fourth component is still emitting. All the other three components have been destroyed (i.e. they concluded their life-cycle) and they start to behave as quenchers or restraints preventing the PL emission. In that range the trend of $\mu_4(t)$ is always decreasing and consequently, the resistances of the system to prevent the PL emission of the corresponding component are going to be lowered. The pumping mechanism dominates the dynamics forcing the excited electrons to change their spin and move to the singlet excited state (the energy gap has been reduced) so that they can relax and finally emit the fluorescence. However, the lowering of the defenses is counterbalanced by the resistances due to radiative friction and the retarding forces acting on the charges during the PL emission. The pumping mechanism is preserved till the reduced mass get its minimum, but from that point on the damping takes over. In fact, in addition to the resistances linked to the emission, the presence of the excited electrons in the singlet excited state activates the damping mechanism, which is the decay pathway governed by the intersystem crossing. As observed in Sections 3 and 4, the intersystem crossing moves the electron to the triplet state impoverishing the excited singlet state and obstructing the PL emission. In fact, after 13,385h, the damping mechanism cannot be neglected anymore and the system starts to rapidly "quench" the luminescence. There are no other relevant components in this time range that are pumping against the damping mechanism produced by all four components. For this reason, the end tail of the normalized PL emission described by a compressed KWW function can be further approximated via a two-term Prony series, see (19). As explained in Section 4, the approximation (19) emerged assuming that $K_D<0$, or in other words we suppose that the pumping mechanism has been suppressed and the damping mechanism dominates the dynamics. Here, the reverse intersystem crossing cannot compete with the triplet-trap rate induced by the quenchers, and we observe the abrupt decay after 13,385h.

Indeed, there are no (relevant) pumping mechanisms after 10,000h. In fact, as for the reference sample, at the end of the PL emission decay only the fourth component is still surviving and when the minimum of the reduced mass has been overcome the decrement in the PL is unstoppable.\\
As observed in Section 5, the compressed KWW function is not complete monotone as the stretched KWW function. According to our model, this property suggests that there will be another reconfiguration whose manifestation is the appearance of a plateau. The reconfiguration is expected and unavoidable since the dual nature of the underlying dynamics implies the existence of a transition from the pumping to the damping that fixes a time window where the luminescence is logarithmically constant (i.e. it behaves ideally and it decays as a purely exponential function), as confirmed by the analysis carried out by the Prony series.  

As observed in Section 6, the end tail of the normalized PL emission is described by a compressed exponential, and so another plateau is being expected. This plateau, which until now had gone unnoticed, appears at $t=40,000$h, but it results to be more evident in the reference sample alq63-3 than in alq65-1. As happened for the previous plateau around 5000h, the correspondence between the reference sample and the annealed one is still ongoing, and this occurrence gives a physical evidence of our approach. In particular, focusing our attention on the zoom in Fig. \ref{fig8}, it is possible to note that the best fit and the Prony series approximation flank the region where the plateau “appears” and give a practical information of its position. Moreover, the time that elapses between the two plateau gives an indirect measurement of the pumping time range needed to populate the singlet excited state.\\
The presence of the plateau at $40,000$h paves the way to the introduction of a fifth component. However, since its intensity is extremely low, approximately ten times less than the corresponding amplitude of the fourth component, it is more convenient to truncate the model to the fourth component. At the end, we can consider the plateau at $4500$h as a transition from an amorphous to crystalline phase transition, or in other words an early stage of the crystalline phase $\alpha$ whereas the last plateau at $40,000$h can be associated with the so-called $\alpha$ phase transition. 

It is possible to draw the same conclusions for the abrupt decay observed in Fig. \ref{fig11} for the annealed sample alq65-1. 
\begin{center}
\includegraphics[scale=0.45]{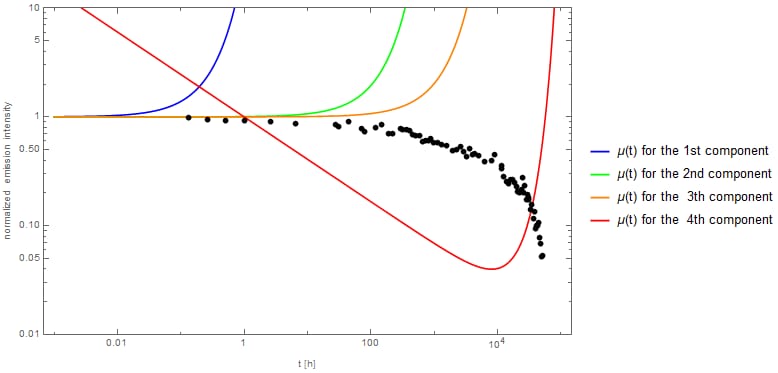}
\captionof{figure}{ PL intensity of the sample alq63-1 decaying in Air (full black circles) annealed at 180$^\circ$C in dry $O_2$ and the time-dependent reduced mass at different relaxation times corresponding to each component: $\tau=0.61$h solid blue line, $\tau=300$h solid green solid, $\tau=2819$h solid orange line and finally $\tau=32,451$h solid red line.}
    \label{fig11}
\end{center}

As a final remark, it is worthwhile to note that the first two components in both samples share some peculiarities. The KWW parameters $\beta_1$ and $\beta_2$ are equal to 1 and the time constants $\tau_1$ and $\tau_2$ differ by two orders in magnitude. Moreover, these early stages of the time-resolved PL emission show the same kind of dynamics. These observations allow us to suppose that the first two components described by pure exponential functions can be merged in one that is stretched as required by the single dynamics. To support this idea, we should consider Table 1 that collects the decay rates (i.e. the inverse of the characteristic time) and the amplitudes of the two-terms Prony series approximating a stretched KWW function. This interpretation allows to emphasize the dynamics governing the first part of the PL emission giving a further physical insight in the quality of the fit. Assuming that the first two components can be merged in only one that is stretched, it follows that the retarding and frictional forces acting on the excited emitters started their action from the very beginning creating micro-environments and configurations that highlight the slow degradation in the time-resolved PL intensity emission.

\section{Summary}\label{sez8}
The relaxation of various physical systems has been found to follow the Kohlrausch-
Williams-Watts (KWW) function, as for example in glassy materials \cite{c14,c22,c23,c24,c25}, in glass-forming liquids \cite{c24} and in the luminescent decay in solid state materials \cite{c1,c2,c3,c4,c5,c6,c7,c8,c9,c10,c11,c13,c16,c17,c18,c19,c20,c26}. Although there is a quantitative agreement between the experimental data and the theoretical fit, the physical meaning of the KWW function is quite elusive. In order to find out the meaning and the origin of the KWW function, a phenomenological model has been introduced and its description has been divided into three steps.

At first, the anomalous relaxation governed by the KWW function has been simplified by choosing an alternative system of reference, called material frame of reference, where the relaxation behaves ideally thanks to the introduction of the so-called material clock \cite{c33,c34,c35,c36,c37,c38}. The material clock marks the time in a different way from the laboratory one: it can slow down or accelerate the passing of time from the pure and mono-exponential behavior that the system experienced during the relaxation. The main result of this step is the definition of a second order differential equation, which describes a damped harmonic oscillator characterized by a time-dependent reduced mass $\mu(t)$ and a time-dependent frequency $\omega(t)$. This differential equation restored the symmetry from a dynamical point of view between the dispersion and the relaxation models governed by the Lorentz model. Analyzing the singularities in the time-dependent coefficient, we were able to understand that the nature of the divergence in the first derivative is only purely mathematical and it is linked to the frame of reference considered. In other words, the origin is a regular singular point. This approach is validated by the fact that the corresponding Hamiltonian generalizes the well-known Caldirola-Kanai Hamiltonian. The introduction of a time-dependent reduced mass $\mu(t)$ in the laboratory frame of reference provides the relaxation dynamics with two distinguishing behaviors, the stretched KWW function and the compressed KWW function on the base of the dynamics of the system. There are two dynamics. A single dynamics based on a damping mechanism and a dual dynamics where there are two mechanisms in series, a pumping mechanism followed by a damping one. The stretched KWW function is governed only by a damping mechanism (a single or monotone dynamics), whereas the compressed KWW function describes a dual dynamics where both these mechanisms occur. In case of dual dynamics, the main feature in the time-resolved photoluminescence is the faster than exponential decay. This gives rise to a plateau in the photoluminescent intensity emission in correspondence of the minimum of the reduced mass $\mu(t)$. This plateau appears when a quasi-equilibrium is established between the pathways followed by the luminophores during their relaxation. This quasi-equilibrium can be considered as a quasi-phase transition.

Then, the analysis is worked out approximating the KWW function via the Prony series. This approximation paves the way to an interesting result: it highlights the meaning of the KWW function as a superposition of the photoluminescent decays. This gives further evidences of the differences between the two KWW functions. The stretched KWW function can be approximated by a sum of simple exponential functions, whereas the compressed KWW function has in its series expansion both positive and negative terms. The presence of the negative terms is due to the fact that the luminophores, i.e. the emitting centers, are quenched after interacting with more active species as the molecular oxygen $O_2$ in the atmosphere.

The final approach consists in the introduction of the mathematical concept of monotonicity and complete monotonicity. These mathematical properties are able to sum up all the physical properties, features and results found in the analysis of the relaxation. They give a physical insight of their importance in modeling relaxation processes.

Among all the physical systems that experience relaxation processes, the photoluminescence from organic molecules as the Tris(8-hydroxyquinoline)aluminum (Alq3) has been considered. The choice of this molecule in the state of thin film as a test-case is based both on the large amount of data available and the widely applications of this material in organic light-emitting diodes (OLEDs), a common technology nowadays. The large amount of data allow us to have a complete overview on the time-resolved PL emission giving the possibility to have enough material to put on a fairy trial the approach proposed.\\ In the future,  we plan to extend and generalize the same approach also to other photoluminescent materials in the vast collection of organic materials and to other photoluminescent functions (as for example in \cite{lattanzi02}). 

\section*{Acknowledgments} 
AL was funded by the Polish National Agency for Academic Exchange NAWA project: Program im. Iwanowskiej PPN/IWA/2018/1/00098 and was supported by the Polish National Center for Science NCN research project OPUS 12 no. UMO-2016/23/B/ST3/01714.



\begin{thebibliography}{99}

\bibitem{c1}Baldacchini G 2020 \textit{Organometallic Luminescence. A Case Study on Alq$_3$, an OLED Reference Material}. Woodhead Publishing.

\bibitem{c2}Baldacchini G, Chiacchiaretta P, Montereali RM, Pode RB, and Vincenti MA 2018 Singular photoluminescence behavior of alq3 lms at very long decay time. \textit{Journal of Luminescence}, {\bf 193}:106-13.

\bibitem{c3}Baldacchini G, Baldacchini T, Chiacchiaretta P, Pode RB, and Wang QM 2009 Morphological phase transitions in alq3 films. \textit{Journal of Luminescence}, 129(12):1831--34.

\bibitem{c4}Baldacchini G, Chiacchiaretta P, Reisfeld R and Zigansky E 2009 The origin of luminescence blueshifts in Alq3 composites.\textit{J. Lumin}, {\bf 129}(12):1849-52.

\bibitem{b}Berberan-Santos MN, Bodunov EN, Valeur B 2005 Mathematical functions for the analysis of luminescence decays with underlying distributions 1. Kohlrausch decay function (stretched exponential), \textit{J. Chem. Phys.} {\bf 315}(1–2)8:171-82.

\bibitem{c5}Bodunov EN, Antonov Y, Simoes A and Gamboa AL 2017 On the origin of stretched exponential (Kohlrausch) relaxation kinetics in the room temperature luminescence decay of colloidal quantum dots. \textit{J. Chem. Phys.}, {\bf 146}(11):114102.

\bibitem{c6}Chen R 2003 Apparent stretched-exponential luminescence decay in crystalline solids. \textit{J. Lumin.}, {\bf 102}:510-8.

\bibitem{c7}Cipelletti L, Pitard E, Weitz DA, Ramos L, Manley S, Pashkovski EE and Johansson M 2003 Universal non-diffusive slow dynamics in aging soft matter, \textit{Faraday discussions}, {\bf 123}:237-51.

\bibitem{c8}Cipelletti L, Trappe V, Bissig H, Romer S and Schurtenberger P 2003 Intermittent dynamics and hyper-aging in dense colloidal gels. \textit{Phys. Chem. Comm.} {\bf 6}(5):21-3.

\bibitem{c9}Cipelletti L, Bissig H, Ballesta P, Trappe V and Mazoyer S 2002 Time-resolved correlation: a new tool for studying temporally heterogeneous dynamics. \textit{J. Phys. Condens. Matter}, {\bf 15}(1):S257.


\bibitem{c10}Dattoli G, G\'orska K, Horzela A, and Penson KA 2014 Photoluminescence decay of silicon nanocrystals and levy stable distributions, \textit{Phys. Lett. A}, {\bf 378} (30-31):2201-5.

\bibitem{c11}Fisher BR, Eisler HJ, Stott NE, and Bawendi MG 2004 Emission intensity dependence and single-exponential behavior in single colloidal quantum dot fluorescence lifetimes. \textit{J. Phys. Chem. B}, {\bf 108}(1):143--48.

\bibitem{c12}Fusco C, Gallo P, Petri A, and Rovere M 2002 Stretched exponential relaxation in a diffusive lattice model. \textit{Phys. Rev. E}, {\bf 65}(2):026127.

\bibitem{c13}Gabriel J, Blochowicz T, and St\"uhn B 2015 Compressed exponential decays in correlation experiments: The infuence of temperature gradients and convection. \textit{J. Chem. Phys.}, {\bf 142}(10):104902.

\bibitem{c14}Gallo P, Rovere M, Ricci MA, Hartnig C and Spohr E 2000 Non-exponential kinetic behaviour of confined water, \textit{E.P.L} {\bf 49}(2):183.

\bibitem{kww04}Garrappa R, Mainardi F, and Maione G 2016 {Models of dielectric relaxation based on completely monotone functions}, \textit{FCAA} {\bf 19}.5:1105--60.

\bibitem{c15}G\'orska K, Horzela A, Dattoli G, Penson K and Duchamp GHE 2017 The stretched exponential behavior and its underlying dynamics. the phenomenological approach. \textit{FCAA} {\bf 20}(1):260--83.

\bibitem{c16}Benny Lee KC, Siegel J, Webb SED, Leveque-Fort S, Cole MJ, Jones R, Dowling K, Lever MJ and French PMW 2001 Application of the stretched exponential function to fluorescence lifetime imaging. \textit{Biophysical J}, {\bf 81}(3):1265--74.

\bibitem{c17}Lee M, Kim J, Tang J, and Hochstrasser RM 2002 Fluorescence quenching and lifetime distributions of single molecules on glass surfaces. \textit{Chem. Phys. Lett.}, {\bf 359}(5-6):412--19.

\bibitem{c18}Madsen A, Sprung M, Leheny RL, Guo H and Czakkel O 2010 Beyond simple exponential correlation functions and equilibrium dynamics in x-ray photon correlation spectroscopy. \textit{New J. Phys.} {\bf 12}(5):055001.

\bibitem{c19}El Masri D, Berthier L, Pierno M and Cipelletti L 2005 Ageing and ultra-slow equilibration in concentrated colloidal hard spheres. \textit{J. Phys. Condens. Matter}, {\bf 17}(45):S3543.

\bibitem{c20}Mazoyer S, Cipelletti L, and Ramos L 2009 Direct-space investigation of the ultraslow ballistic dynamics of a soft glass. \textit{Phys. Rev. E}, {\bf 79}(1):011501.

\bibitem{c21}Milovanov AV, Rypdal K and Rasmussen JJ 2007 Stretched exponential relax-ation and ac universality in disordered dielectrics. \textit{Phys. Rev. B} {\bf 76}(10):104201.

\bibitem{ngai}Ngai KL 2011 Relaxation and Diffusion in Complex Systems, Springer, New York.

\bibitem{c22}Phillips JC 1996 Stretched exponential relaxation in molecular and electronic glasses. \textit{Rep. Prog. Phys.} {\bf 59}(9):1133.

\bibitem{c23}Phillips JC 1995 Kohlrausch relaxation and glass transitions in experiment and in molecular dynamics simulations, \textit{J. Non-Crist. Solid.} {\bf 182}(1-2):155--61.

\bibitem{c24}Xia X and Wolynes PG 2001 Microscopic theory of heterogeneity and non exponential relaxations in supercooled liquids. \textit{Phys. Rev. Lett.} {\bf 86}(24):5526.

\bibitem{c25}G\"otze W 2008 \textit{Complex dynamics of glass-forming liquids: A mode-coupling theory}, {\bf 143} OUP Oxford.

\bibitem{c26}Schlegel G, Bohnenberger J, Potapova I and Mews A 2002 Fluorescence decay time of single semiconductor nanocrystals, \textit{Phys. Rev. Lett.} {\bf 88}(13):137401.

\bibitem{kww01}Kohlrausch R 1854 Theorie des elektrischen Rückstandes in der Leidener Flasche. Ann. Phys. {\bf 167}(2):179--214.

\bibitem{kww02}Williams G and Watts DC 1970 Non-Symmetrical Dielectric
Relaxation Behavior Arising from a Simple Empirical Decay Function. \textit{Trans. Faraday Soc.} {\bf 66}:80–-5.

\bibitem{kww03}Anderssen RS, Husain SA, and Loy RJ 2003 \textit{The Kohlrausch function: properties and applications}. Anziam J {\bf 45}:C800-C816.


\bibitem{lukichev}Lukichev A 2019 \emph{Physical meaning of the stretched exponential Kohlrausch function}. \textit{Phys. Lett. A} {\bf 383}(24): 2983--87.

\bibitem{c27}Barnes D 2013 LCD or OLED: Who wins? \textit{SID Symposium Digest of Technical Papers}, {\bf 44}:26--7. Wiley Online Library.

\bibitem{c28}Buckley A 2013 Organic light-emitting diodes (OLEDs): materials, devices and applications. Elsevier.

\bibitem{c29}Geffroy B, Le Roy P and Prat C. 2006 Organic light-emitting diode (OLED) technology: materials, devices and display technologies. \textit{Polym. Int.}, {\bf 55}(6):572--82.

\bibitem{c30}Maier SA 2007 \textit{Plasmonics: fundamentals and applications}. Springer Science $\&$ Business Media.

\bibitem{c31}Schols S 2011 \textit{Device Architecture and Materials for Organic Light-Emitting Devices: Targeting High Current Densities and Control of the Triplet Concentration}. Springer Science $\&$ Business Media.

\bibitem{c32}Tang CW and VanSlyke SA 1987 Organic electroluminescent diodes. \textit{Applied physics letters}, {\bf 51}(12):913--15.

\bibitem{c33}Findley WN and Davis FA 2013 \textit{Creep and relaxation of nonlinear viscoelastic materials}. Courier Corporation.

\bibitem{c34}Hecksher T, Olsen NB, Niss K and Dyre JC 2010 Physical aging of molecular glasses studied by a device allowing for rapid thermal equilibration \textit{J. Chem. Phys.}, {\bf 133}(17):174514.

\bibitem{c35}Martin R 2008 \textit{Ageing of composites}. Elsevier.


\bibitem{c36}Merodio J and Ogden RW 2020 \textit{Constitutive Modelling of Solid Continua}. Number 262. Springer.


\bibitem{c37}Narayanaswamy O 1971 A model of structural relaxation in glass. \textit{J. Amer. Cer. Soc.}, 54(10):491--98.

\bibitem{c38}Roth CB 2016 \textit{Polymer glasses}. CRC Press.

\bibitem{bib23}Bender CM and Orszag SA 2013 Advanced mathematical methods for scientists and engineers I: Asymptotic methods and perturbation theory, Springer Science $\&$Business Media.

\bibitem{bib90}Humi M and Miller W 2012 Second course in ordinary differential equations for scientists and engineers, Springer Science $\&$ Business Media.

\bibitem{bib108}Kristensson G 2010 Second order differential equations: special functions and their classification, Springer Science $\&$ Business Media.

\bibitem{bib154}Rainville ED 1958 Elementary differential equations-2.

\bibitem{bib155}Redheffer RM and Port D 1991 Differential equations: theory and applications. Jones $\&$ Bartlett Learning.

\bibitem{c39}Caldirola P 1941 Forze non conservative nella meccanica quantistica. \textit{Il Nuovo Cimento}, {\bf 18}(9):393--400.

\bibitem{c40}Kanai E 1948 On the quantization of the dissipative systems. \textit{Prog. Theor. Phys.}, {\bf 3}(4):440--42.

\bibitem{hpaper}Huang MC and Wu MC 1998 The Caldirola-Kanai model and its equivalent theories for a damped harmonic oscillator, \textit{Chin. J. Phys.}, {\bf 36}(4): 566--87.

\bibitem{c41}Doss K, Yang Y, Wilkinson CJ, Huang L, Lee KH and Mauro JC 2020 Maxwell relaxation time for non-exponential relaxation phenomena in glassy systems. \textit{J. Amer. Cer. Soc.}.

\bibitem{c42}Zheng Z,  Mauro JC and Allan D 2018 Modeling of delayed elasticity in glass. \textit{J. Non-Cryst. Sol.}, {\bf 500}:432--42.

\bibitem{c51}Penson KA and G\'orska K 2010 Exact and explicit probability densities for one-sided levy stable distributions. \textit{Phys. Rev. Lett.}, {\bf 105}(21):210604.


\bibitem{c52}G\'orska K and Penson KA 2011 Levy stable two-sided distributions: Exact and explicit densities for asymmetric case, \textit{Phys. Rev. E}, {\bf 83}(6):061125.

\bibitem{c43}Engel T 2006 \textit{Physical chemistry}. Pearson Education India.


\bibitem{c44}Baleiz{\~a}o C and Berberan-Santos MN 2007 Thermally activated delayed fluorescence as a cycling process between excited singlet and triplet states: Application to the fullerenes. \textit{J. Chem. Phys.}, {\bf 126}(20):204510.

\bibitem{c45}Fukumura H, Kazuhide K, Kikuchi K and Kokubun H 1988 Temperature effect on inverse intersystem crossing of anthracenes. \textit{J. Photch. Photobio. A}, {\bf 42}(2-3):283--91.

\bibitem{c46}Tanaka F, Okamoto M, and Hirayama S 1995 Pressure and temperature dependences of the rate constant for s1-t2 intersystem crossing of anthracene compounds in solution. \textit{J. Phys. Chem.}, {\bf 99}(2):525--30.

\bibitem{c47}Miller KS and Samko SG 2001 Completely monotonic functions. \textit{Integr. Transf. Spec. F.}, 12(4):389--402.



\bibitem{c48}Bernstein S et al. 1929 Sur les fonctions absolument monotones. \textit{Acta Math.}, {\bf 52}:1-66.


\bibitem{c49}Song R, Schilling RL and Vondracek Z 2012 Bernstein functions: theory and applications, {\bf 37}, Walter de Gruyter.


\bibitem{c50}Pollard H 1946 The representation of $e^{-x^{\lambda}}$ as a Laplace integral. \textit{Bull. Amer. Math. Soc.}, 52(10):908--10.

\bibitem{das}Das S 2017 Revisiting the Curie-von Schweidler law for dielectric relaxation and derivation of distribution function for relaxation rates as Zipf’s power law and manifestation of fractional differential equation for capacitor, \textit{J. Mod. Phys.} {\bf 8}.12: 1988-2012.

\bibitem{c53}Bergstrom H 1952 On some expansions of stable distribution functions. \textit{Arkiv for Matematik}, 2(4):375--78.


\bibitem{lattanzi02}Lattanzi A, Casasanta G, Garra R 2021 On the application of Mittag-Leffler functions to hyperbolic-type decay of luminescence, ArXiv


\end{thebibliography}
\end{document}